\documentclass[pra,preprint]{revtex4} 
\usepackage{silence}
\usepackage{amsmath,revsymb,graphicx,amssymb,youngtab}

\newcommand{\kt}[1]{\ensuremath{|#1\rangle}}
\newcommand{\br}[1]{\ensuremath{\langle#1|}}
\newcommand{\bk}[2]{\ensuremath{\langle #1|#2\rangle}}
\newcommand{\HS}{\mathcal{H}}
\newcommand{\SHS}{\mathcal{S}}
\newcommand{\KHS}{\mathcal{K}}

\begin{document}

\title{Symmetry and Natural Quantum Structures\\
 for Three-Particles in One-Dimension}

\author{N.L.~Harshman\footnote{Electronic address: harshman@american.edu}}

\affiliation{Department of Physics\\
4400 Massachusetts Ave.\ NW\\ American University\\ Washington, DC 20016-8058}

\begin{abstract}
How do symmetries induce natural and useful quantum structures? This question is investigated in the context of models of three interacting particles in one-dimension. Such models display a wide spectrum of possibilities for dynamical systems, from integrability to hard chaos. This article demonstrates that the related but distinct notions of integrability, separability, and solvability 
identify meaningful collective observables for Hamiltonians with sufficient symmetry.
In turn, these observables induce tensor product structures on the Hilbert space that are especially  useful for storing and processing quantum information and provide guidance to interpretation and phenomenology of quantum few-body physics.
\end{abstract}

\maketitle

\section{Introduction}

This article presents a collection of examples demonstrating how quantum structures are relevant to the study of three particles in one dimension. The collection is curated around a theme: symmetry selects preferred quantum structures, and preferred quantum structures instantiate observation, information, control and interpretation `naturally'. The ironic quotes around `naturally' indicate that this may be an idiosyncratic (and possibly recursive) definition. Whatever `naturally' means, physicists know that some representations, coordinate systems, reference frames, etc.\ are more natural than others for studying specific systems. While underlying physics must be independent of representation, a judicious choice of representation simplifies analysis by eliminating irrelevant variables, indicating conserved quantities, consolidating dimensional variables into unified scales, allowing approximation methods to converge more rapidly, and so on.
Often finding the `better' representation requires a transformation from observables associated with reductive constituents to observables associated with the emergent collective. So a thread running through the collection is how symmetry identifies collective observables that induce physically-meaningful quantum structures for the three-body system. 

Why three bodies? Three-body systems have been at the frontier of dynamical analysis since Newton, and their study is the first baby step from two-body to many-body systems. The three-body problem is perennially fertile; new periodic solutions to the even classical three-body gravitational system (in three dimensions) have been discovered recently~\cite{suvakov}. This article considers models with three non-relativistic particles in one dimension.  Even in one dimension, the general three-body problem is not integrable and does not have exact and entire solutions unless some additional symmetries provide extra integrals of the motion~\cite{chencinier}. When additional symmetry is not present, some classical, non-integrable three body-systems have soft chaos~\cite{gutzwiller}: they stay close to integrable models where perturbation theory works well, at least for certain initial conditions. Other models have hard chaos and deterministic prediction is impractical for realistic uncertainties and any initial conditions.

Diagnosing how integrability affects the emergence of `natural' collective quantum structures is a central (and unresolved) goal of this inquiry. Integrability means there are as many independent integrals of the motion as degrees of freedom, and these integrals of the motion are observables in involution (defined by the Poisson bracket) with the Hamiltonian and with each other. When a system is classically integrable, the integrals of motion certainly provide a `natural' way to describe the system in terms of action-angle variables. The complete set of integrals are determined by initial conditions and define a manifold in phase space that is essentially a multi-dimensional torus. The trajectory lies on the manifold and it is trivially expressed in terms of action-angle coordinates, although the canonical map back to position and momentum coordinates may be highly non-trivial. Note that integrable is not the same thing as separable in the classical sense, i.e.\ explicitly finding a set of coordinates that separate the Hamilton-Jacobi equation so that the equations of motion can be solved by quadrature. 
Separability is used in this article both in the sense of separable coordinates for solving the three-body Schr\"{o}dinger equation and separable from the quantum information perspective of Hilbert space tensor product structures and entanglement. Clarifying that connection is another goal.

There is still debate about how best to carry over the definition of integrability to quantum systems~\cite{weigert, caux,braak,larson}. Some aspects of integrability transition neatly from classical mechanics: integrals of the motion are realized as operators on the Hilbert space that commute with the Hamiltonian and each other. However, sticking points include how to extend to discrete degrees of freedom and finite Hilbert spaces, how to define independence of integral operators, and how to deal with indistinguishable particles. 
For certain three-body scenarios, where integrability is clear, a reasonable hypothesis is that the set of commuting observables associated to integrals of motion induce tensor product structures that are `natural'. For example, when quantum structures are induced by observables that commute with the Hamiltonian, information measures based on those structures, like entanglement, are invariant in time. And sometimes these structures indicate preferred collective degrees of freedom that are compatible with symmetrization of identical particles.
Using a broad definition of information, by choosing quantum structures that align with integrability, we find a representation for the system in which the embodied information can be shifted to the observables for `typical' states.

Another motivation for this study is recent experiments with ultracold atoms in effectively one-dimensional optical traps. In the lab, the two-body interactions between the atoms can be tuned by Feshbach and confinement-induced resonances and are well-approximated by delta-function contact potential~\cite{Olshanii1998}.  
See \cite{cazalilla} and \cite{guan} for reviews of theory and experiment for one-dimensional bosonic and fermionic systems, respectively, and \cite{harshman2015a,harshman2015b} for an overview of symmetries of the few-body model and recent theoretical references. One interesting feature of this model is that the strength of the contact interaction is described by a single parameter, and as this parameter varies from zero to infinity the model  interpolates between two integrable cases, the non-interacting limit and the so-called unitary limit.  
Although conclusive statements are not yet forthcoming, this system demonstrates how `natural' quantum structures fade and re-emerge as symmetry and integrability is tuned between two solvable models.

This introduction has argued that one-dimensional, quantum three-body models are a minimally-complicated but experimentally-relevant system to investigate open questions about the relationships among integrability, separability, and symmetry. The rest of the article is a series of essays on these relationships and how they induce quantum structures. Thinking about quantum structures, and in particular the relativity of entanglement~\cite{zanardi01, zanardi04, barnum04, harshman11, dugic}, has proved useful for a variety of purposes, including developing decoherence-free subspaces for error-correcting codes~\cite{viola01}, understanding the dynamics of open systems~\cite{arsenijevic}, and classifying entanglement structures in two-body systems~\cite{harshman12}. 
The goal here is to see whether thinking about quantum structures can possibly explain the emergence of `natural' observables for storing and processing quantum information. Since the models under consideration are relevant to ultracold atomic gases, one of the possible working materials for proposed quantum information devices like computers and simulators, this inquiry into quantum structures has practical ramifications.  A hypothesis is that systems near solvable models will work best in such applications, and so the first section reviews some examples of solvable models used throughout the article.

\section{Solvable Models}

The family of models under consideration have a Hamiltonian $\hat{H}$ expressed in terms of the particle position and momentum operators $\hat{X}_i$ and $\hat{P}_i$ like
\begin{equation}\label{hambasic}
\hat{H} =  \sum_i \left( \frac{1}{2m} \hat{P}_i^2 + V^1(\hat{X}_i) \right) + \sum_{\langle i, j \rangle} V^2(|\hat{X}_i - \hat{X}_j|),
\end{equation}
where the second sum is over all pairs ${\langle i, j \rangle}$. The one-particle external potential $V^1(x)$ ($x\in\mathbb{R}$) and the two-particle interaction potential $V^2(r)$ ($r\in\mathbb{R}^+$) are assumed to be sufficiently well-behaved functions so that when the position observables $\hat{X}_i$ are inserted into them, the resulting operators are still self-adjoint on (a dense domain of) the Hilbert space. Notice that I am only considering spin-independent Hamiltonians in this article. Although not dynamically coupled to spatial degrees of freedom, spin degrees of freedom still play a role due to symmetrization of identical indistinguishable particles.

For certain choices of the functions $V^1$ and $V^2$, the Hamiltonian (\ref{hambasic}) is solvable. Solvable models are like lamps illuminating the landscape of dynamical systems. They are touchstones for analysis because (as the joke goes~\cite{calogero1971}) if you are looking for your lost keys, under the lampposts is where there is enough light for you to search effectively. 

\subsection{Solvability for One Particle}

First, note that all one-particle, one-dimensional systems are integrable. The one-particle Hamiltonian
\[
\hat{h} = \frac{1}{2m} \hat{P}^2 + V^1(\hat{X})
\]
certainly commutes with itself, so there are as many integrals of motion (one) as there are degrees of freedom (also one). Classically, the equation of motion can be solved by quadrature for the classically allowed spectrum of energies. Quantum mechanically, the energy levels of bound states (the only kind of states considered here) are discretized by a quantum number $n\in\{0,1,2,\ldots\}\equiv\mathbb{N}$ that labels energies $\epsilon_n$ in the one-particle  spectrum $\{\epsilon_0,\epsilon_1,\epsilon_2,\ldots\} = \sigma^1$. 
In principle, each $\epsilon_n$ is the solution of an integral equation that depends on the functional form of $V^1$, so in addition to being integrable, all one-dimensional, one-particle systems in an external potential are also solvable.

When the spectrum $\sigma^1$ is known, one can always define a function $\epsilon(z)$ by a finite series such that $\epsilon(n) = \epsilon_n$ for all $n \leq n_{max}$, with $n_{max}$ some arbitrary cutoff. When the function $\epsilon(n)$ is a finite algebraic expression in terms of the quantum number $n$ for all $n\in\mathbb{N}$, the system is algebraically solvable. Algebraically solvable one-dimensional potentials include the infinite square well, harmonic oscillator, P\"{o}schl-Teller, and Morse potentials.

\subsection{Non-interacting Three-Particle Models}

When there are no interactions $V^2 \equiv 0$, every $\hat{H}$ with form (\ref{hambasic}) is integrable because $\hat{H} = \hat{h}_1 +\hat{h}_2 + \hat{h}_3$ decomposes into a sum of three independent one-particle, one-dimensional Hamiltonians $\hat{h}_i$. The total non-interacting spectrum $\sigma_0$ is all possible sums of three elements of $\sigma^1$, and therefore all energies $E\in\sigma_0$ are
\begin{itemize}
\item non-degenerate when $E = 3 \epsilon_j$;
\item three-fold degenerate when $E = 2 \epsilon_j + \epsilon_k$ ($j\neq k$);
\item six-fold degenerate when $E = \epsilon_j + \epsilon_k + \epsilon_l$ ($j\neq k \neq l$).
\end{itemize}

Harmonic traps $V^1(x_i) = m\omega x_i^2/2$ are an example of a three-particle system that is superintegrable, i.e.\ the Hamiltonian has more independent integrals of motion than the three degrees of freedom~\cite{evans}. The three-particle model is isomorphic to one particle in an isotropic harmonic potential in three-dimensions with nine independent invariant operators~\cite{jauch,louck,kalnins}: the three single particle Hamiltonians $\hat{h}_i$, the three `angular momentum'-like  operators $\hat{Q}_i\hat{P}_j - \hat{P}_i \hat{Q}_j$, and the Demkov operators $\hat{P}_i\hat{P}_j + m^2 \omega^2 \hat{X}_i \hat{X}_j$.

\subsection{Interacting Three-Particle Models}

When interactions are included, here are three notable solvable cases:
\begin{itemize}
\item Models with harmonic interactions $V^2(|x_i - x_j|)=\gamma (x_i - x_j)^2$ are solvable when the external field is quadratic $V^1(x_i) = A x_i^2 + B x_i + C$. In particular, for harmonic traps ($A>0$) and homogeneous potentials ($A = B=0$) the model is algebraically solvable. This solvability is a consequence of the separability of the center-of-mass and relative degrees of freedom for quadratic potentials; any other external potential function $V^1(x)$ does not preserve this separation. Further, for harmonic traps, the interacting system is superintegrable because an angular-momentum-like operator in remains invariant. 
\item Models with the contact interaction $V^2(|x_i - x_j|)=\gamma \delta(x_i - x_j)$ are solvable in the unitary limit $\gamma \rightarrow \infty$, and are algebraically solvable whenever the one-particle potential $V^1(x)$ is algebraically solvable. Further, for the case of a homogeneous potential and infinitely hard-wall boundary conditions, the model is solvable for any value of $\gamma$ by the technique of Bethe's ansatz~\cite{oelkers}. This case is an example where integrability does not follow from some obvious choice of separable coordinates.
\item The Calogero-Moser model with the inverse-square potential $V^2(|x_i - x_j|)=\gamma /(x_i - x_j)^2$ and the harmonic trap potential $V^1(x_i) = m \omega^2 x_i^2/2$ is algebraically solvable~\cite{calogero1971}. There are also several other related solvable models when $V^2(|x_i - x_j|)$ is a reciprocal sinusoidal or hyperbolic function~\cite{calogero2008}.
\end{itemize}

Ideally, there would be an unified method to characterize separability, integrability, solvability and algebraic solvability in terms of symmetries. And certainly, much work has been done here, but no complete, coherent picture has emerged. However, for integrable and solvable systems, the integrals give `natural' observables that induce `natural' quantum structures on the Hilbert space that are useful for analyzing and extracting entanglement, as the following sections explore.

\section{Separability in Configuration Space}

Let us identify one chain of separability, a more refined notion than integrability, that selects a preferred tensor product structure on the Hilbert space. 
\begin{enumerate}
\item Identify a natural Hamiltonian that is totally separable. In other words, there is a canonical coordinate transformation that separates the Hamilton-Jacobi equation into a sum of action-angle contributions.
\item The quantum system is also separable in that coordinate system, i.e.\ the configuration space representation of the Hamiltonian is a separable partial differential equation. For each spatial degree of freedom there is a countably-infinite tower of harmonics. Products of these harmonics are energy eigenstates with a phase that simply rotates in time. 
\item There are specific observables associated with the separable coordinate system. These observables are constructed from the generators of the symmetries implied by the separability of the system. Sometimes the generators for the algebra are inherited from an underlying model with more symmetry.  The algebra of observables can be partitioned further into commuting subalgebras for each independent degree of freedom. 
\item These subalgebras induce a partitioning of the Hilbert space. In other words, the tensor products of the representations of the subalgebras are the same as the representation of the entire system. Since these observables commute with the Hamiltonian, entanglement with respect to this tensor product structure is invariant with respect to time.
\end{enumerate}
This sounds like a plausible method for inducing a `natural' tensor product structure for classically separable systems, but the devil is in the details. This section teases out a few of those details, and for the limited set of models considered here, arrives at a heuristic classification.

One advantage of working with three identical particles in one dimension is that the three-dimensional configuration space has  familiar geometrical structures, coordinates and transformations. One realization of the spatial Hilbert space $\KHS$ is the space of Lebesgue-square-integrable functions $L^2(\mathcal{X})$ on the configuration space $\mathcal{X} \sim \mathbb{R}^3$. The manifold $\mathcal{X}$ has many different possible coordinate systems. The simplest case is where each coordinate is just the particle position $ x_i \in (-\infty, \infty)$ with respect to a common origin in $\mathcal{X}$, corresponding to the decomposition $\mathbb{R}^3 = \mathbb{R} \times \mathbb{R} \times\mathbb{R}$. 
This is just one of an equivalence class of rectangular coordinate systems for $\mathcal{X}$ related by translation, rotation and orthogonal scaling.

There are a host of other coordinate systems, but in physics the most interesting from the perspective of quantum structures and solvability for the three-body, one-dimension problem are the orthonormal coordinate systems that separate the Schr\"{o}dinger equation in three dimensions. Delightfully, these have already been entirely classified. Depending on the functional form of the potential $V({\bf x})$, the time-independent Schr\"{o}dinger equation 
\begin{equation}\label{hamcoord}
-\frac{\hbar^2}{2 m}\left( \frac{\partial^2}{\partial x_1^2} + \frac{\partial^2}{\partial x_2^2}  + \frac{\partial^2}{\partial x_3^2} \right)\Psi({\bf x}) + \left( V({\bf x})-E \right)\Psi({\bf x}) = 0.
\end{equation}
can be solved by separation of variables in eleven different orthonormal coordinate systems~\cite{eisenhart,morse}: rectangular, cylindrical, elliptical cylindrical, parabolic cylindrical, spherical, conical, parabolic, prolate spheroidal, oblate spheroidal, ellipsoidal and paraboloidal. All of these coordinate systems can be unified into a single framework, i.e.\ they are all degenerations of ellipsoidal coordinates. For each coordinate system, the transformed potential $\tilde{V}(\xi)$ in the separable coordinates $\xi=\{\xi_1,\xi_2,\xi_3\}$ must have a particular form in order for (\ref{hamcoord}) to separate. 
For example, in spherical coordinates $\{r,\theta,\phi\}$ the potential must have the form
\[
\tilde{V}(\xi) = v_1(r) + \frac{v_2(\theta)}{r^2} + \frac{v_3(\phi)}{r^2 \sin^2 \theta}.
\]
For our case of three identical particles interacting via Galilean-invariant interactions, the potential $V({\bf x})$ is constrained to the form
\begin{equation}\label{v3part}
V({\bf x}) = V^1(x_1)+ V^1(x_2)+ V^1(x_3) + V^2(|x_1-x_2|) + V^2(|x_2-x_3|)+ V^2(|x_3-x_1|)
\end{equation}
where $V^1(x_i)$ is the external trap potential and $V^2(|x_i-x_j|)$ is the Galilean invariant two-body interaction between particles $i$ and $j$. The specific form (\ref{v3part}) eliminates most of the eleven possibilities for separable potentials. When there are non-trivial interactions $V^2$, only rectangular and cylindrical remain, and then only for harmonic traps. 
(Harmonic traps with no interactions are separable in eight of the eleven coordinate systems~\cite{kalnins}.)

Even when the model Hamiltonian is separable, there are still distinctions between different kinds of separability depending on the nature of the separation constants~\cite{morse}. This has consequences for the quantum structures that can be defined, and I categorize these into three standards of separability: gold, silver and bronze.

\subsection{Gold Separability}

The most straightforward example of separability of the Schr\"{o}dinger equation leading to separable degrees of freedom is rectangular coordinates, when some linear transformation of the particle coordinates ${\bf q} = R{\bf x} + {\bf b}$ separates the potential $V({\bf x})$ into a sum of independent functions like 
\[
\tilde{V}({\bf q}) = v_1(q_1) + v_2(q_2) + v_3(q_3).
\]
When there are are no two-body interactions, every three-body system is certainly of this form with $R=I_3$ and $q_i=x_i$, but there are interacting examples, too. Quadratic potentials $V^1(x_i) = A x_i^2 + B x_i + C$   with harmonic two body interactions $V^2(|x_i - x_j|) = \gamma(x_i - x_j)^2$ are separable in rectangular Jacobi rectangular coordinates:
\begin{eqnarray}\label{jacrec}
q_1 &=& \frac{1}{\sqrt{3}}\left( x_1 +x_2 +x_3 \right)\nonumber\\
q_2& =& \frac{1}{\sqrt{2}}\left( x_1 - x_2  \right)\nonumber\\
q_3 &=& \frac{1}{\sqrt{6}}\left( x_1 + x_2 - 2 x_3 \right).
\end{eqnarray}
Note that any orthogonal combination $q'_2 = \cos\vartheta q_2 + \sin\vartheta q_3$ and $q'_3 = -\sin\vartheta q_2 + \cos\vartheta q_3$ (where $\vartheta\in [0,2\pi)$) also separates the quadratic trap with harmonic interactions.

Why is this the gold standard? Because systems separable in rectangular coordinates correspond most directly to our notions of separable degrees of freedom and independent modes, and therewith to a `naturally' separable Hilbert space. The Hamiltonian breaks into a sum of three terms $H= h_1 + h_2 +h_3$ where
\[
h_i = -\frac{\hbar^2}{2 m} \frac{\partial^2}{\partial q_i^2} + v_i(q_i).
\]
The energy $E = \epsilon_1(\eta_1) + \epsilon_2(\eta_2) + \epsilon_3(\eta_3)$ can be expressed as a sum of functions $\epsilon_i(\eta_i)$, each of which is a function of only one of the three separation constants $\{\eta_1,\eta_2,\eta_3\}$.  The energy eigenstates are products of functions $X_i$ which each depend on single separation constant $\eta_i$ as
\[
X_1(q_1;\eta_1)X_2(q_2;\eta_2)X_3(q_3;\eta_3).
\]
Denote the spectrum for each separation constant $\eta_i$ by $\sigma_i$. These spectra $\sigma_i$ are independent for rectangular coordinates and so the general state $\kt{\Phi}\in\KHS$ is expressed as an independent triple sum over a basis labeled by the separation constants (i.e.\ quantum numbers) $\eta_i$
\[
\kt{\Phi} = \sum_{\eta_1\in\sigma_1}  \sum_{\eta_2\in\sigma_2} \sum_{\eta_3\in\sigma_3} c_{\eta_1\eta_2\eta_3} \kt{\eta_1\eta_2\eta_3}.
\]
Put another way, each subspace can be treated mathematically like a totally independent subsystem.  The spatial Hilbert space is the tensor product
\begin{equation}\label{rectsep}
\KHS= \KHS_1 \otimes \KHS_2 \otimes \KHS_3
\end{equation}
where each $\KHS_i$ is realized by square-summable sequences $\ell^2(\eta_i)$ labeled by the quantum number $\eta_i$. The abstract Hamiltonian $\hat{H}$ (as opposed to its representation $H$ on $\mathcal{X}$) is the sum of three local operators
\begin{equation}\label{sumform}
\hat{H} = \hat{h}_1 \otimes \hat{\mathbb{I}} \otimes \hat{\mathbb{I}} + \hat{\mathbb{I}} \otimes \hat{h}_2 \otimes \hat{\mathbb{I}} + \hat{\mathbb{I}} \otimes \hat{\mathbb{I}} \otimes \hat{h}_3.
\end{equation}
The set of observables $\{\hat{h}_1,\hat{h}_2,\hat{h}_3\}$ is complete and commuting, so by Zanardi's theorem~\cite{zanardi01,zanardi04,harshman11} these subalgebras of observables induce the tensor product structure (\ref{rectsep}) on $\KHS$.
This relatively simple example of separability in rectangular coordinates provides a useful contrast with the cases below.

\subsection{Silver and Bronze Separability}

Unlike gold separability, for general separable coordinate systems $\xi = \{\xi_1,\xi_2,\xi_3\}$, the different form of the Hamiltonian cannot by written as a sum of three local sub-Hamiltonians. Although  the eigenstates are separable functions of the coordinates, they are not separable with respect to the separation constants $\eta_i$  as
\[
X_1(\xi_1;\eta_1,\eta_2,\eta_3)X_2(\xi_2;\eta_1,\eta_2,\eta_3)X_3(\xi_3;\eta_1,\eta_2,\eta_3).
\]
What distinguishes silver and bronze is whether the spectra $\sigma_i$ of separation constants  are independent (silver) or not (bronze). 

As an example consider two cases of silver separability in Jacobi cylindrical coordinates $\{\rho,\phi,z\}$:  
 the harmonic trap with either harmonic interactions
\begin{eqnarray}\label{harmharm}
H &=& -\frac{\hbar^2}{2 m} \nabla^2 + \frac{1}{2} m \omega^2 {\bf x}^2 + \gamma\left((x_1-x_2)^2 + (x_2-x_3)^2 + (x_3-x_1)^2 \right)\nonumber\\
&=& -\frac{\hbar^2}{2 m} \nabla^2 + \frac{1}{2} m \omega^2 q_1^2 + \left( \frac{1}{2} m \omega^2 + 3 \gamma  \right)\rho^2
\end{eqnarray}
or the Calogero-Moser Hamiltonian with inverse quadratic interactions
\begin{eqnarray}\label{cm}
H &=& -\frac{\hbar^2}{2 m} \nabla^2 + \frac{1}{2} m \omega^2 {\bf x}^2 + \frac{\gamma}{(x_1-x_2)^2} + \frac{\gamma}{(x_2-x_3)^2} + \frac{\gamma}{(x_3-x_1)^2}\nonumber\\
&=& -\frac{\hbar^2}{2 m} \nabla^2 + \frac{1}{2} m \omega^2 (q_1^2 + \rho^2) + \frac{2 \gamma}{\rho^2}\left( \frac{1}{\cos^2\phi} + \frac{1}{\cos^2(\phi-2\pi/3)}+ \frac{1}{\cos^2(\phi-4\pi/3)}\right)\nonumber\\
&=& -\frac{\hbar^2}{2 m} \nabla^2 + \frac{1}{2} m \omega^2 (q_1^2 + \rho^2)+ \frac{18 \gamma}{\rho^2 \sec^2(3\phi)},
\end{eqnarray}
where $\rho^2 = q_2^2+q_3^2$ and $\tan\phi = q_3/q_2$ in terms of Jacobi rectangular coordinates (\ref{jacrec}).

In both of these cases the potential in Jacobi cylindrical coordinates satisfies
\[
\tilde{V}(\rho,\phi,z) = v_\rho(\rho) + \frac{v_\phi(\phi)}{\rho^2} + v_z(z).
\]
However, note that the configuration space Hamiltonian cannot be written as a sum of three differential operators that each only depend on a single variable. The second term ties together the separation constants $\eta_\rho \equiv \nu$ and $\eta_\phi \equiv \mu$ (but not $\eta_z\equiv\eta$) and the general energy eigenstate has the form
\[
R_{\nu\mu}(\rho)\Phi_\mu(\phi)Z_{\eta}(z).
\]
In principle one could imagine defining a tensor product structure like
\[
\KHS_\rho \otimes \KHS_\phi \otimes \KHS_z \sim L^2(\mathbb{R}^+) \otimes L^2(\mathbb{S}_1) \otimes L^2(\mathbb{R}),
\]
but the Hamiltonian would not have a decomposition like (\ref{sumform}). Therefore, this is not `gold' separability.

Although the separation constants are not independent, their spectra are, and that is what makes both of theses cases `silver' separable in this classification scheme. In terms of the separation constants, the energy for the harmonic interaction Hamiltonian (\ref{harmharm}) is
\[
E_{\eta\nu\mu} = \hbar\omega (\eta + 1/2) + \hbar\sqrt{\omega^2 + 4 m \gamma} (2\nu + |\mu| +1)
\]
with $\nu\in\sigma_\nu\mathbb{N}$, $\eta\in\sigma_\eta= \mathbb{N}$ and $\mu\in\sum_\mu = \mathbb{Z}$. For the Calogero-Moser Hamiltonian (\ref{cm}) the energy is
\[
E_{\eta\nu\mu} = \hbar\omega \left[\eta + 2\nu + |\mu|+ 3/2\left( 1 + \sqrt{1 + 2m^2 \gamma} \right) \right]
\]
again with $\nu\in\sigma_\nu =\mathbb{N}$ and $\eta\in\sigma_\eta= \mathbb{N}$, but this time only for $\mu/3\in\mathbb{Z}$, i.e.\ $\sigma_\mu$ contains all integer multiples of $3$. Because the spectra of separation constants are independent,
for both cases the states can still be expressed in terms of energy eigenstates like
\[
\kt{\Phi} = \sum_{\nu\in\sigma_\nu}  \sum_{\mu\in\sigma_\mu} \sum_{\eta\in\sigma_\eta} c_{\nu\mu\eta} \kt{\nu\mu\eta}
\]
where the order of the sum does not matter.

An example of `bronze' separability is the harmonic, non-interacting model solved in spherical coordinates $\{r,\theta,\phi\}$ with the usual separation constants $\{n,l,m\}$. The energy eigenstates have the form 
\[
R(r;n,l)\Theta(\theta;l,m)\Phi(\phi;m),
\]
and that is  definitely not gold separability because of the multiple separation constant dependence of the polar and radial harmonics. Further, the spectrum of possible $m$ depends on $l$, so the triple sum
\[
\kt{\Phi} = \sum_{n\in\mathbb{N}}  \sum_{l \in \mathbb{N}} \sum_{m=-l}^l c_{nlm} \kt{nlm}
\]
cannot be rearranged arbitrarily~\footnote{For the familiar Coulomb potential (which has no equivalent one-dimensional three-body model), $\sigma_l$ also depends on $n$ and so the separability is even `bronzer' than the isotropic harmonic trap.}.

Why do these distinctions among different types of separability matter? A quantum structure is useful from the point of view of control and information when there are experimentally-accessible observables that are complete and local with respect to that tensor product structure. For silver and bronze separability, the Hamiltonian does not decompose into a sum of local subsystem Hamiltonians on a partitioned Hilbert space, so the virtual subsystems induced by such a partition are less `natural'. However, at least for the silver separability there are complete and local operators that characterize the spectra and basis for each partition, making this case  somewhat more natural. For example, it seems more physically reasonable to imagine controllably entangling the radial and polar degrees of freedom for the two cases in cylindrical coordinates than controlling entanglement between the $m$-degree of freedom and the $l$-degree of freedom in spherical coordinates. 
Future work will try to make this fuzzy notion of `reasonable' and `natural' well-defined, and the key will likely be better understanding of the relationship between separability and symmetry.

\section{Types of Symmetries}

Here I classify some different types of symmetries that generate structures in the Hilbert space via their representations. Of most concern are configuration space and phase space symmetry transformation groups that leave the Hamiltonian invariant, but dynamical symmetries are briefly mentioned at the end because of their connection to separability. 

\subsection{Configuration Space Symmetries}

Repeating for convenience, in configuration space ${\bf x} = \{x_1, x_2,x_3\}\in \mathcal{X}$ the configuration space Hamiltonian $H$ has the form
\begin{equation}\label{hamconfig}
H = -\frac{\hbar^2}{2m} \nabla^2 + \sum_i V^1(x_i) + \sum_{\langle i, j \rangle} V^2(|x_i - x_j|).
\end{equation}
Here I restrict attention to cases where $V^1(x_i)$ is a one-body trapping potential such that $\lim_{x\rightarrow \infty} V^1(\pm x) \rightarrow \infty$ and $V^2(|x_i - x_j|)$ is a Galilean-invariant two-body interaction potential.

A configuration space symmetry is a group of transformations of $\mathcal{X}$. These transformations do not have to be linear, but of present interest are subgroups of orthogonal transformations $O\in\mathrm{O}(3)$. An orthogonal transformation $O$ on configuration space induces a unitary transformation $\hat{U}(O)$ on the Hilbert space by
\begin{equation}
\psi(O {\bf x}) = \bk{O{\bf x}}{\psi} = \br{{\bf x}}\hat{U}^\dag(O)\kt{\psi} = (\hat{U}(O^{-1})\psi)({\bf x}).
\end{equation}
When there is a complete basis for the Hilbert space, this method can be used to explicitly calculate the representation (although this is usually not the most practical method).

For otherwise arbitrary functions $V^1$ and $V^2$, the only symmetry that the Hamiltonian $H$ necessarily has in configuration space is the geometrical realization of particle permutation symmetry $\mathrm{P}_3$, which is isomorphic to the abstract symmetric group on three objects $\mathrm{S}_3$. The six elements of $\mathrm{P}_3$ can broken into the following:
\begin{itemize}
\item The identity, which maps $\{123\}$ into $\{123\}$ and acts as the identity matrix $I_3$ on $\mathcal{X}$. 
\item The pairwise exchanges $\{213\}$, $\{132\}$ and $\{321\}$ which act as reflections in $\mathcal{X}$ across the planes $x_1 = x_2$, $x_2 = x_3$ and $x_1 = x_3$, respectively.
\item The 3-cycles $\{231\}$ and $\{312\}$ that act as rotations by $\pm 2\pi/3$ around the line $x_1 = x_2 = x_3$.
\end{itemize}
This set of six transformations is isomorphic to the three-dimensional point group denoted $\mathrm{C}_{3v}$ in Sch\"{o}nflies notation or $[3]$ in Coxeter notation.

Notice that all six of these transformations are orthogonal transformations $O \in\mathrm{O}(3)$ that leave invariant the Jacobi cylindrical coordinates $q_1$ and $\rho$. The quantity $\rho$ is expressed in particle coordinates like
\[
\rho^2 = \sum_{\langle i, j \rangle} (x_i - x_j)^2 = 2x_1^2 + 2x_2^2 + 2x_3^2 - 2x_1x_2 -2x_2x_3 - 2x_1x_3
\]
and it quantifies the scale of the relative particle distribution, i.e.\ the Euclidean distance in $\mathcal{X}$ from the line of three-particle coincidence $x_1 = x_2 = x_3$.

The Hamiltonian $H$ can have additional configuration space symmetries if the trap is parity symmetric $V^1(-x+ 2b ) = V^1(x)$ about some point $b$. Denote by $\Pi_i$ the configuration space operator realizing this reflection $x_i \rightarrow -x_i + 2b$. A single particle inversion $\Pi_i$ does not leave the Hamiltonian invariant because of the interaction terms~\footnote{In the absence of interactions, each single particle parity $\Pi_i$ is a symmetry of $H$, but the $\Pi_i$ do not commute with particle permutations. The group structure that describes that combination is the wreath product (a kind of semi-direct product) of the symmetric group with the reflection group $ \mathrm{Z}_2 \wr \mathrm{S}_3$.}. However, the total inversion $\Pi = \Pi_1 \Pi_2 \Pi_3$ does leave $H$ invariant, and it commutes with all elements of $\mathrm{P}_3$. 
Putting this together, the minimal configuration space symmetry of $H$ is $\mathrm{P}_3 \times \mathrm{O}(1) \sim \mathrm{S}_3 \times \mathrm{Z}_2$, where the inversion subgroup is denoted $\mathrm{O}(1)$. This group is isomorphic to the point group $D_{3d} \sim [[3]]$.

For certain traps and potentials, there is more than the minimal configuration space symmetry for three interacting particles $\mathrm{P}_3$ (no parity) or $\mathrm{P}_3 \times \mathrm{O}(1)$ (parity). Clearly there is more than minimal symmetry in a harmonic well with any interaction potential:
\begin{eqnarray}\label{hamharm}
H &=&-\frac{\hbar^2}{2m} \nabla^2 + \frac{m \omega^2}{2} (x_1^2 + x_2^2 + x_3^2) + V^2(|x_1-x_2|) +  V^2(|x_2-x_3|) + V^2(|x_1-x_3|) \nonumber\\
 &=& -\frac{\hbar^2}{2m} \nabla^2 + \frac{m \omega^2}{2} (\rho^2 + q_1^2) \nonumber\\
&& {} + V^2(\rho|\cos\phi|) +  V^2(\rho|\cos(\phi-2\pi/3)|) + V^2(\rho|\cos(\phi-4\pi/3)|).
\end{eqnarray}
In Jacobi cylindrical coordinates, total inversion takes the form $\Pi\cdot\{q_1, \rho,\phi\} = \{-q_1, \rho,\phi + \pi\}$. As mentioned above, the rotations and reflection in $\mathrm{P}_3$ leave $q_1$ and $\rho$ invariant but transform $\phi$. For example, the particle exchange $\{213\}$ maps $\phi$ into $\pi -\phi$ and the 3-cycle $\{231\}$ maps $\phi$ to $\phi+2\pi/3$. But now there is an additional independent symmetry: the relative and center-of-mass degrees of freedom can be independently inverted. 
In other words, relative parity $\Pi_r\cdot\{q_1, \rho,\phi\} = \{q_1, \rho,\phi + \pi\}$ is a good quantum number for Hamiltonians with harmonic traps. 
This means that for three particles in a harmonic trap with arbitrary two-particle interactions the configuration space symmetry group is $\mathrm{P}_3\times\mathrm{O}(1) \times \mathrm{O}(1)$ and is isomorphic to $\mathrm{D}_{6h} \sim [[3],2]$. This additional symmetry is an example of an emergent symmetry, i.e.\ a symmetry of the model than cannot be generated by products of one-particle operators like $\Pi_i$ and permutations in $\mathrm{P}_3$. 

A group of non-linear symmetry transformations in configuration space are relevant for the Calogero-Moser Hamiltonian (\ref{cm}) for any interaction strength or for unitary limit $\gamma\rightarrow\infty$ of the contact interaction Hamiltonian in any external trap
\begin{equation}\label{unitarycontact}
H =-\frac{\hbar^2}{2m} \nabla^2 + \sum_i V^1(x_i) + \gamma\left(\delta(x_1-x_2) +  \delta(x_2-x_3) + \delta(x_1-x_3)\right).
\end{equation}
For these models, position wave functions with finite energy must vanish on the coincidence manifold $\mathcal{V}$ defined as the union of the planes $x_1=x_2$, $x_2 = x_3$ and $x_1 = x_3$. The effective configuration space is therefore $\mathcal{X}' = \mathcal{X} - \mathcal{V}$. The manifold $\mathcal{V}$ divides $\mathcal{X}$ into six equivalent sectors $\mathcal{X}_{ijk}$, each with fixed particle order $x_i > x_j > x_k$. The Hilbert space of finite-energy states $\KHS' \subset \KHS$ is realized by
\begin{equation}\label{sector}
L^2(\mathcal{X}') \sim L^2(\mathcal{X}_{ijk}) \otimes \mathbb{C}^6.
\end{equation}
In other words, every eigenstate of (\ref{cm}) or (\ref{unitarycontact}) is six-fold degenerate for distinguishable particles. Sector permutations act as orthogonal transformations of the degeneracy space, and the solvability of those two models can be understood as a consequence of this symmetry. A permutation of the six sectors $\mathcal{X}_{ijk}$ is a non-linear transformation of $\mathcal{X}'$ because of the discontinuity across the coincidence planes.
 For the contact interaction, the energy eigenfunctions in each sector $\mathcal{X}_{ijk}$ are just the fermionic (totally-antisymmetric) solutions to the non-interacting problem with the same trapping potential $V^1$ restricted to the sector, a result usually attributed to Girardeau~\cite{girardeau}.

\subsection{Phase Space Symmetries}

Any orthogonal transformation $O$ of configuration space $ {\bf q} = O {\bf x}$ induces an orthogonal, canonical transformation in six-dimensional phase space $\{ {\bf q}, {\bf  k}\} = \{ O{\bf x}, O{\bf  p}\}$ that is also a symmetry transformation of $\hat{H}$. There is always at least one additional symmetry in phase space beyond configuration space, namely time translation $\mathrm{T}_t$ generated by $\hat{H}$ itself. Classically, this is just the Hamiltonian flow in time along the trajectories through phase space. The quantum mechanical consequence for a completely bound system is the discretization of energy.

So for an asymmetric trap with no additional symmetries, the phase space symmetry group is $\mathrm{T}_t \times \mathrm{P}_3$ and for a symmetric trap it is $\mathrm{T}_t \times \mathrm{P}_3 \times \mathrm{O}(1)$. As before, for a harmonic trap like (\ref{hamharm}), there is an additional, emergent symmetry due to the separable center-of-mass degree of freedom. Any rotation in the $q_1$-$k_1$ hyperplane, where $k_1$ is the momentum conjugate to $q_1$, is a phase space symmetry. 
This is the standard $\mathrm{U}(1)$ symmetry of the one-dimensional harmonic oscillator. The group $\mathrm{U}(1)$ contains the relative parity operation, so the total phase space symmetry for three-harmonically trapped particles is $\mathrm{T}_t \times \mathrm{P}_3 \times \mathrm{O}(1) \times \mathrm{U}(1)$.

\subsection{Dynamic Symmetries}

More generally, the group of all unitary operators that commute with the Hamiltonian is the kinematic symmetry group of the Hamiltonian. Only kinematic symmetry groups associated to transformations of the configuration space or phase space have been considered; the case of accidental kinematic symmetries is briefly discussed in the next section.
Sometimes there are transformations of configuration space, phase space, or the Hilbert space that do not commute with the Hamiltonian but map its energy spectrum onto itself in a regular way. These are examples of dynamic symmetries, also known as spectrum-generating symmetries. 
 In a Lie-algebraic dynamic symmetry, the Hamiltonian is not invariant under a group of transformations of the phase space, but the generators of that transformation combined with the Hamiltonian form a Lie algebra of operators defined on the Hilbert space. One example of a Lie-algebraic dynamic symmetry is the `hidden' $\mathrm{SO}(2,1)$ symmetry of the harmonic oscillator~\cite{pitaevskii,werner}. First consider the one-particle case and define the operators
\begin{equation}
\hat{W}^\pm = \frac{1}{4\sqrt{2}} \left( \frac{1}{m\hbar\omega} \hat{P}^2  -\frac{m\omega}{\hbar} \hat{X}^2 + \frac{i}{\hbar} \hat{X}\hat{P} + \frac{i}{\hbar} \hat{P}\hat{X}\right).
\end{equation}
These operators have the commutation relation with the one-particle harmonic Hamiltonian $\hat{h}$
\[
[\hat{h}, \hat{W}^\pm] = \pm 2 \hbar \omega \hat{W}^\pm,
\]
showing that they act like ladder operators connecting states with an energy difference of $2 \hbar \omega$, i.e.\ two rungs apart. The commutator between the ladder operators is
\[
[\hat{W}^+,\hat{W}^-] = -\frac{1}{2 \hbar\omega}\hat{h}
\]
where the minus sign indicates that this an $\mathrm{SO}(2,1)$-like Lie algebra and not an $\mathrm{SO}(3)$-like Lie algebra.

This dynamic $\mathrm{SO}(2,1)$ symmetry  can be extended to three non-interacting particles in a harmonic trap in several ways and it is related to the separability of radial coordinates (which have a length scale) and angular coordinates (which do not)~\cite{werner}. This dynamic symmetry-induced separability can also be extended to interacting Hamiltonians that separate in Jacobi cylindrical coordinates like harmonic trap/harmonic interaction Hamiltonian (\ref{harmharm}), the Calogero-Moser Hamiltonian (\ref{cm}), and the unitary limit of the contact interaction Hamiltonian (\ref{unitarycontact}).

For an example of a different kind of dynamic symmetry relevant for few-body systems called state permutation symmetry~\cite{Chen}, see \cite{harshman2015a,harshman2015b}.

\section{Symmetry and Structures}

There is a natural tensor product structure for systems with three particles with spin in one dimension:
\begin{equation}\label{Htripart}
\HS = \HS_1 \otimes \HS_2 \otimes \HS_3.
\end{equation}
This tensor product structure can be induced by the one-particle subalgebras of observables.
Further, since the one-particle spin operators commute with the spatial operators, they form complete, commuting subalgebras and each particle's Hilbert space can be factored into a spatial part $\KHS_i$ and a spin part $\SHS_i$ like
\begin{equation}\label{spatspin1}
\HS_i = \KHS_i \otimes \SHS_i \sim \mathrm{L}^2(\mathbb{R}) \otimes \mathbb{C}^{J},
\end{equation}
where $J$ is the number of spin (or internal) discrete levels for each particle. 

This article only considers spin-independent Hamiltonians, so the spaces $\SHS_i$ are only relevant for symmetrization of identical particles. Therefore, the six-fold tensor product structure implied by the composition of (\ref{Htripart}) and (\ref{spatspin1}) can be repartitioned into the dynamically-invariant tensor product structure
\begin{equation}\label{spatspin3}
\HS = \KHS \otimes \SHS \sim \mathrm{L}^2(\mathbb{R}^3) \otimes \mathbb{C}^{J^3},
\end{equation}
where
\[
\KHS=\bigotimes_i \KHS_i\ \mbox{and}\ \SHS=\bigotimes_i \SHS_i.
\]
The one-parameter group of time translation $\mathrm{T}_t$ with elements $t \in \mathbb{R}$ is represented by the unitary operator $\hat{U}(t)$ on $\HS$ which factors on the tensor product (\ref{spatspin3})
\[
\hat{U}(t) = \hat{U}_\KHS(t) \otimes \hat{\mathbb{I}},
\]
i.e.\ it is local with respect to the structure (\ref{spatspin3}). One way to think about this is that the spin-independent Hamiltonian $\hat{H}$ has $\mathrm{U}(J^3)$ symmetry since it commutes with any unitary operator acting only on $\SHS$. The dynamical-invariance of the tensor product structure (\ref{spatspin3}) means that the amount of entanglement between the spatial degrees of freedom and spin degrees of freedom is an invariant in time. For example, any pure initial state $\kt{\Psi(0)}$ can be decomposed like
\begin{equation}
\kt{\Psi(0)} = \sum_j \lambda_j \kt{\kappa_j(0)} \otimes \kt{\sigma_j}
\end{equation}
with orthogonal vectors $\kt{\kappa_j(0)} \in \KHS$ and $\kt{\sigma_j} \in \SHS$. Although the spatial states $\kt{\kappa_j(t)}$ evolve in time, the coefficients $\lambda_j$ that measure the entanglement between spin and spatial degrees of freedom do not.

The connections among the spin-invariance of the Hamiltonian, commuting subalgebras of observables, the induced `natural' tensor product structure, and the consequences for entanglement dynamics are reasonably straightforward. The rest of this section develops some more formal notions of how symmetry representations induce quantum structures and then gives a few examples. 

\subsection{Irreducible Representation Decompositions}

Symmetries induce quantum structures through their representation on the Hilbert space. Denote a kinematic symmetry group by $\mathrm{G}$. In the article, only finite or compact groups appear as kinematic groups, so all irreducible representations (irreps) of $\mathrm{G}$ can be realized by unitary finite-dimensional matrices acting on an irrep space $\mathcal{M}^\mu$.  Denote these irreps by $D^{\mu}$ where $\mu$ labels the irrep. The dimension of the irrep is $d(\mu)$ and the irrep space is $\mathcal{M}^{\mu} \sim \mathbb{C}^{d(\mu)}$. The Hilbert space can be decomposed into subspaces called irrep towers $\mathcal{H}^{\mu}$
\begin{equation}\label{hilbertdecomp1}
\HS = \bigoplus_{\{\mu\}} \mathcal{H}^{\mu}.
\end{equation}
The set of irreps $\{\mu\}$ for a finite group is finite, and for a general compact group it is countable. Each irrep tower $\mathcal{H}^{\mu}$ can be decomposed into a direct sum of equivalent irrep spaces $\mathcal{M}^{\mu}_i \sim \mathbb{C}^{d(\mu)}$ like
\begin{equation}
\mathcal{H}^{\mu} = \bigoplus_i \mathcal{M}^{\mu}_i.
\end{equation}
This decomposition of the tower $\mathcal{H}^{\mu}$ is formally isomorphic to the direct product of a single irrep space $\mathcal{M}^\mu$ with a space $\mathcal{A}^\mu$ that represents the degeneracy
\begin{equation}\label{towerdecomp2}
\bigoplus_i \mathcal{M}^{\mu}_i = \mathcal{M}^\mu \otimes  \mathcal{A}^\mu.
\end{equation}
The irrep degeneracy space $\mathcal{A}^\mu$ can be finite or infinite dimensional. The tensor product decomposition (\ref{towerdecomp2}) is `natural' when the space $\mathcal{A}^\mu$ is diagonalized by an observable $\hat{A}_\mu$ (or set of observables) that is defined on the tower $\mathcal{H}^\mu$, that commutes with $G$, and that distinguishes different appearances of the same irrep $\mathcal{M}^\mu$. Such an observable $\hat{A}_\mu$ induces a tensor product structure that `separates' the irrep tower $\mathcal{H}^{\mu}$.

Better from the point of view of `natural' separability is when the same observable $\hat{A}$ is defined over all $\HS$ and it separates all the towers. Then the space $\mathcal{A}$ is the same for all towers and can be factored out of the decomposition like
\begin{equation}\label{hilbertdecomp2}
\HS = \left( \bigoplus_{\{\mu\}} \mathcal{M}^\mu \right) \otimes  \mathcal{A}.
\end{equation}
Depending on the operator (or set of operators) $\hat{A}$, the space $\mathcal{A}$ also may carry a representation of another symmetry group, say $\mathrm{G}'$. If so, then the process can be repeated: $\mathcal{A}$ can be decomposed into $\mathrm{G}'$-towers, perhaps inducing another separable factor.

\subsection{Example of Symmetry-Induced Structure I: Time Translation}

Let us bring this back to the case of three trapped particles. The most important symmetry group has already been mentioned, time-translation symmetry $\mathrm{T}_t$. The representation on the Hilbert space is of course just $\hat{U}(t) = \exp(-i\hat{H}t/\hbar)$. The group $\mathrm{T}_t \sim \mathbb{R}$ is abelian and has one-dimensional irreps $\mathcal{M}^E$ characterized by a real number, the energy $E$. The Hilbert space is therefore decomposable into irrep towers
\begin{equation}
\HS = \bigoplus_{E \in\sigma} \mathcal{H}^E,
\end{equation}
where $\sigma$ is the spectrum of eigenvalues of $\hat{H}$ on $\HS$.
If there were no other symmetries, then the space $\mathcal{H}^E$ is one-dimensional and isomorphic to the $\mathrm{T}_t$-irrep $\mathcal{H}^E\sim\mathcal{M}^E$. More generally, each energy level has a degeneracy $d(E)$ and is decomposed as
\[
\mathcal{H}^E = \bigoplus_{i=1}^{d(E)} \mathcal{M}^E = \mathcal{A}^E \sim \mathbb{C}^{d(E)}.
\]
Note that any operator that commutes with the Hamiltonian decomposes into a block diagonal form, acting only on the spaces $\mathcal{A}^E$. The goal of spectroscopy (from the limited perspective of a group theorist) is therefore to find the symmetry group $\mathrm{G}$ of the Hamiltonian and its representation on the whole Hilbert space $\HS$ such that every degenerate energy subspace $\mathcal{A}^E$ is an irrep space $\mathcal{M}^\mu$ of $\mathrm{G}$. When that is possible, there is a unique association from every $E \in\sigma$ onto some irrep $\mu$ of the group $\mathrm{G}$. Then $\HS$ has the alternate decomposition
\begin{subequations}\label{nodegen}
\begin{equation}
\HS = \bigoplus_{\{\mu \}} \mathcal{H}^\mu
\end{equation}
and
\begin{equation}
\mathcal{H}^\mu =\bigoplus_{E \in \sigma|_\mu} \mathcal{M}^\mu_E = \mathcal{M}^\mu \otimes \mathcal{A}^E
\end{equation}
\end{subequations}
where the labels for the copies of $\mathcal{M}_E^\mu$ into which the tower $\mathcal{M}^\mu$ decomposes are in fact just the energies $E$ in the spectrum $\sigma|_\mu$ restricted to only $\mu$-types irreps. For three trapped particles, the spectrum $\sigma|_\mu$ is discrete, bounded from below, and countable, and $\mathcal{A}^E\sim \ell^2(E)$, the square-summable sequences. Whether this kind of partition of the Hilbert space is useful depends very much on the interplay between the spectrum of energies and spectrum of irreps of $\mathrm{G}$.

Again, this decomposition (\ref{nodegen}) applies in the ideal case where the maximal symmetry group of $\hat{H}$ has been identified. If there are energy levels with degeneracies $d(E)$ that do not correspond to irreps dimensions $d(\mu)$, then there are two possibilities: (1) a global symmetry is missing from $\mathrm{G}$, where by `global symmetry' I mean symmetry that is is induced on the Hilbert space from a symmetry transformation defined on the whole configuration space or phase space; or (2) a truly `accidental' symmetry, i.e.\ a conspiracy of parameters that leads to certain energy levels lining up by `accident'. 
The canonical example of this accidental symmetry is would be the Pythagorean degeneracy of certain non-interacting three-particle energy levels when the confining potential is a perfect, infinite cube.

\subsection{Example of Symmetry-Induced Structure II: Particle Permutations}

To make this more concrete, at a minimum the symmetry group of the Hamiltonian must have the subgroup $\mathrm{P}_3 \sim \mathrm{S}_3$. This group has three irreps:
\begin{itemize}
\item The one-dimensional totally symmetric irrep is denoted $[3]$ or $\tiny\yng(3)$. On $\mathcal{M}^{[3]}$, every element $p\in\mathrm{P}_3$ is represented by $1$.
\item The one-dimensional totally antisymmetric irrep is denoted $[1^3]$ or $\tiny\yng(1,1,1)$. On $\mathcal{M}^{[1^3]}$, odd permutations are represented by $-1$ and even permutations by $1$.
\item The two-dimensional irrep with mixed symmetry is denoted $[21]$ or $\tiny\yng(2,1)$. On $\mathcal{M}^{[21]}$, elements $p$ can be represented by orthogonal $2 \times 2$ matrices.
\end{itemize}
Thus, the Hilbert space can be decomposed like
\begin{equation}
\HS = \mathcal{H}^{[3]} \oplus \mathcal{H}^{[21]} \oplus \mathcal{H}^{[1^3]}
\end{equation}
and the states of bosonic identical particles are elements of $\mathcal{H}^{[3]}$ with spectrum $\sigma|_{[3]}$ and fermionic particles are $\mathcal{H}^{[1^3]}$ with spectrum $\sigma|_{[1^3]}$. Consider the case when the particles have no spin or internal components ($J=1$) and so $\HS = \KHS$ and $\SHS$ is trivial. If there are no other symmetries, then each one of these spaces $\HS^\mu$ is a tower of $\mathrm{S}_3$ irreps $\mathcal{M}^\mu$ with different energies. For example, since $\mathcal{M}^{[3]}$ is trivial, the bosonic sector is
\[
\HS^{[3]} = \bigoplus_{E \in \sigma|_{[3]}} \mathcal{M}^{[3]}_{E} 
\]
if there are no other symmetries. In other words, $\HS^{[3]}$ is isomorphic to the space square-summable series $\ell^2(E)$ on the index $E \in\sigma|_{[3]}$. If there are additional symmetries, then there may be degenerate bosonic levels and the decomposition is 
\begin{equation}\label{3sum}
\HS^{[3]} = \bigoplus_{E \in \sigma|_{[3]}} \mathcal{M}^{[3]}_{E} \otimes \mathcal{A}^{E}.
\end{equation}
Then the goal is to find observables that diagonalize the finite-dimensional space $\mathcal{A}^{[3],E}$. Ideally, these spaces (and the  observables that diagonalize them) are the same for every $E$ so an $\dim(\mathcal{A}^{[3]})$-dimensional degeneracy space $\mathcal{A}^{[3],E} \equiv \mathcal{A}^{[3]}$ factors out of the sum (\ref{3sum}) so the irrep tower $\mathcal{H}^{[3]}$ can be realized by $\ell^2(E)\otimes  \mathbb{C}^{\dim(\mathcal{A}^{[3]})}$. This seems to be possible whenever the three particle model is integrable, although a proof has not yet been found.

\section{Conclusion}

The reader who has made it this far deserves two confessions. First, obviously this research is still a work in progress. The goal of finding how symmetry selects preferred quantum structure has not been met and a unifying framework for identifying emergent collective observables has not been found. Nor has a novel practical protocol for manipulating robust and error-protected  quantum information and entanglement based on those quantum structures been discovered. However, pursuing the relationships among integrability, solvability and separability has revealed that not all degrees of freedom are created equal, and that quantum structures induced by some symmetries and observables are more natural than others.

Second, the reader deserves to know a `secret' motivation for this work. Each month it seems another physicist  adds their voice to the `it from bit' chorus, and I admit I find their songs compelling. Perhaps quantum information theory will simultaneously open doorways to the next generation of technology, provide new techniques to resolve difficult quantum problems in everything from superconductivity to nuclear structure, and resolve long-standing interpretational questions.
That would certainly be exciting! But I still do not understand how to define information in quantum theory without conceding some sort of meta-theoretical status to a privileged observer. And so this essay on three particles in one dimension is part of a continuing inquiry into how `nature' can select `natural' observables that turn data into information, and eventually, into physical meaning, without relying on an observer.

\end{document}